\documentclass[twocolumn,preprintnumbers,amsmath,amssymb]{revtex4}
\usepackage{times}
\usepackage{graphicx}
\renewcommand{\d}{{\rm{d}}}

\newcommand{\dfon}[2]{%
\frac{\delta{#1}}{\delta{#2}}}

\begin{document}
\draft
\preprint{xxx}
\title{Functional sensitivity of Burgers and related 
equations to initial conditions}
\author{O. Vall\'ee and E. Moreau}
\address{LASEP UPRES-EA 3269\\
Facult\'e des Sciences -- Universit\'e d'Orl\'eans, \\
rue Gaston Berger BP 4043,\\
18028 Bourges cedex, FRANCE}

\date{\today}

\begin{abstract}
In this paper, we apply sensitivity methods to nonlinear PDEs like 
Burgers and KPZ equations. These equations are known to have 
analytical solutions which make easier the analysis of the 
sensitivity of their solutions to initial conditions. The main result
stands in the fact that the most  the solution is sensitive to the
initial condition, the most it is decorrelated in space, \textit{i.e.} the 
values of the initial condition participate to the solution at all 
distances of the wave front. This finally reveals a particular aspect 
of the Burgers turbulence.
\end{abstract}

\pacs{XXX?YYY,ZZZ}
\maketitle

\section{Introduction}
Burgers and related equations have been introduced in many fields of 
sciences such as non-equilibrium statistical physics. For 
instance, we can cite, in cosmology, the model known as the adhesion model 
\cite{cosmo}, there is also the modeling of traffic jam 
\cite{jam}, the description of 
directed polymers in random media \cite{newman,medina}
or the dynamics of growing interface \cite{kpza}. Note also that the 
Burgers equation may be a draft modeling to fluid dynamics.

When modeling a system by means of the Burgers or the KPZ equation 
knowledge of the initial condition is required. However a great insight 
of the problem is necessary in order to fit experimental data as soon 
as high quality results are available. It is then interesting to know 
the interdependence of the solution of the modeling equation to
initial conditions. This is the goal of sensitivity analysis to provide 
the system response to variations of input \cite{volt,fran}. This 
mathematical method has been applied in numerous domains of sciences, 
see for instance \cite{rab,hang}.

For an initial value 
problem, such as the following evolution equation:
\begin{eqnarray}
   \partial_{t}u(x,t)= A\,u(x,t)\quad\nonumber\\
   u(x,0)=\varphi(x),\;x\in\mathbb{R},
\end{eqnarray}
where $A$ is some nonlinear operator, the variation $\delta\varphi(x)$ 
of the initial condition should imply variations $\delta u(x,t)$ in the solution 
$u(x,t)$ of the equation. These variations obey the functional 
relation:
\begin{equation}
    \delta u(x,t)=\int_{-\infty}^{+\infty}\dfon{u(x,t)}{\varphi(y)}\,
    \delta\varphi(y)\,\d y.
\end{equation}
Then the functional derivative $\delta{u(x,t)}/\delta{\varphi(y)}$ 
(sometimes called the density) gives a 
quantitative measure of the 
response of the actual solution $u(x,t)$ to any variation of the input.

As said extensively in the literature, nonlinear and chaotic systems 
are mainly characterized by sensitivity to initial conditions. The 
purpose of the present paper is to quantify this sensitivity in the 
cases of Burgers and related equations.
\section{The heat equation}
Before we treat the case of nonlinear equation, we shall examine 
in this section the case of the heat equation:
\begin{equation}
    \partial_{t}Z=\nu\partial_{xx}Z,
    \label{cha}
\end{equation}
with the initial condition: 
$Z(x,0)=\psi(x)$, and where $\nu$ is some diffusion parameter. 
We have the well known solution to the heat equation 
as the convolution integral:
\begin{equation}
    Z(x,t)=\frac{1}{\sqrt{4\pi\nu t}}\int_{-\infty}^{+\infty}
    \psi(y)\,\exp\left[-\frac{(x-y)^{2}}{4\nu t}\right]\,\d y.
    \label{sol}
\end{equation}
Now we can calculate the sensitivity coefficient, 
\textit{i.e.} the functional 
derivative of the solution $Z(x,t)$ with respect to the initial condition
$\psi(x)$ directly from this solution. It reads:
\begin{equation}
    \dfon{Z(x,t)}{\psi(x')}=\frac{1}{\sqrt{4\pi\nu t}}
    \int_{-\infty}^{+\infty}
     \dfon{\psi(y)}{\psi(x')}\,\exp\left[-\frac{(x-y)^{2}}{4\nu t}\right]\,\d y.
\end{equation}
As we have: $\dfon{\psi(y)}{\psi(x')}=\delta(x'-y)$, the sensitivity 
coefficient is then given by:
\begin{equation}
    \dfon{Z(x,t)}{\psi(x')}=\frac{1}{\sqrt{4\pi\nu t}}
    \exp\left[-\frac{(x-x')^{2}}{4\nu t}\right].
\end{equation}
We first remark that the density is well correlated along the line 
$x=x'$ in the $(x,x')$ plane and it does not involve the 
initial condition $\psi(x)$ itself. In fact, since the heat equation is 
linear, this density is also a solution to the heat 
equation with the initial condition $\delta(x-x')$ (\textit{i.e.} the fundamental 
solution).
Moreover, as $t$ goes to infinity the sensitivity coefficient spreads 
and goes to zero. Consequently, the solution to the heat 
equation is asymptotically insensitive to initial conditions. 
This proves, as well, the  known issue concerning the very high difficulty
to find the initial condition from the measurement of $Z(x,t)$ at any 
time.
\section{The unforced Burgers equation}
Let us recall, the chain of transformations leading to the solution of 
the unforced Burgers equation.
The Burgers equation is given by:
\begin{equation}
    \partial_{t}u+u\partial_{x}u=\nu \partial_{xx}u,
\end{equation}
with the initial condition $u(x,0)=\varphi(x)$. Where $\nu$ stands 
usually for the viscosity coefficient. The change of 
function defined by $u(x,t)=-\partial_{x}h(x,t)$ leads to the KPZ 
equation \cite{kpza}:
\begin{equation}
    \partial_{t}h=\frac{1}{2}\left(\partial_{x}h\right)^{2}+\nu \partial_{xx}h,
    \label{kpz}
\end{equation}
with the corresponding initial condition $h(x,0)=\eta(x)$. Then the 
Hopf-Cole transformation: $h=2\nu \ln Z$, yields to the heat equation 
(Eq. (\ref{cha})) \cite{kpza,Hopf,cole}.

From the general solution of the heat equation (Eq. (\ref{sol})), we 
find the solution to Eq.(\ref{kpz}) as:
\begin{eqnarray}
   h(x,t)=2\nu\ln\left\{\frac{1}{\sqrt{4\pi\nu t}}
   \qquad\qquad\qquad\qquad\qquad\qquad\right.\nonumber\\
\left.\quad\qquad\times\int_{-\infty}^{+\infty}
    \exp\left[\frac{1}{2\nu}\left(\eta(y)-
    \frac{(x-y)^{2}}{2 t}\right)\right]\,\d y\right\}.
    \label{kpza}
\end{eqnarray}
Taking the functional derivative of this equation, we obtain the 
sensitivity coefficient of $h(x,t)$ to the initial condition:
\begin{equation}
    \dfon{h(x,t)}{\eta(x')}=\frac{\exp\left[\frac{1}{2\nu}\left(\eta(x')-
    \frac{(x-x')^{2}}{2 t}\right)\right]}{\int_{-\infty}^{+\infty}
    \exp\left[\frac{1}{2\nu}\left(\eta(y)-
    \frac{(x-y)^{2}}{2 t}\right)\right]\,\d y}.
    \label{senKPZ}
\end{equation}

On the contrary to the heat equation, the unforced KPZ equation 
solution is 
very sensitive to the initial condition, moreover this is also the 
case in the long time limit:
\begin{equation}
    \lim_{t\to \infty}\dfon{h(x,t)}{\eta(x')}=
    \frac{\exp\left[\frac{1}{2\nu} \eta(x')\right]}
    {\int_{-\infty}^{+\infty}
    \exp\left[\frac{1}{2\nu}\eta(y)\right]\,\d y},
    \label{decor}
\end{equation}
where we assume the convergence of the integral. Surprisingly it 
does not depend on $x$.
This means that, in the 
long time limit, the initial condition $\eta(x')$ at the distance $x'$ influences  
$h(x,t)$ for all the values of $x$ with the same weight.
It is interesting also to look at 
the inviscid limit of this result. Assuming there is only one 
stationary point $a$, the sensitivity coefficient may written as:
\begin{equation}
    \dfon{h(x,t)}{\eta(x')}
    {\xrightarrow{t\to\infty,\,\nu\to 0}}
    \sqrt{\frac{\vert h''(a)\vert}{4\pi\nu}}
    \exp\left[\frac{1}{2\nu}(\eta(x')-\eta(a))\right].
\end{equation}
Note these two limits do not act always uniformly.
As $a$ corresponds to the maximum of $\eta(x)$, we see that the sensitivity 
coefficient is peaked on this value.

\begin{figure}[!ht]
	 \centerline{\includegraphics[scale=0.55]{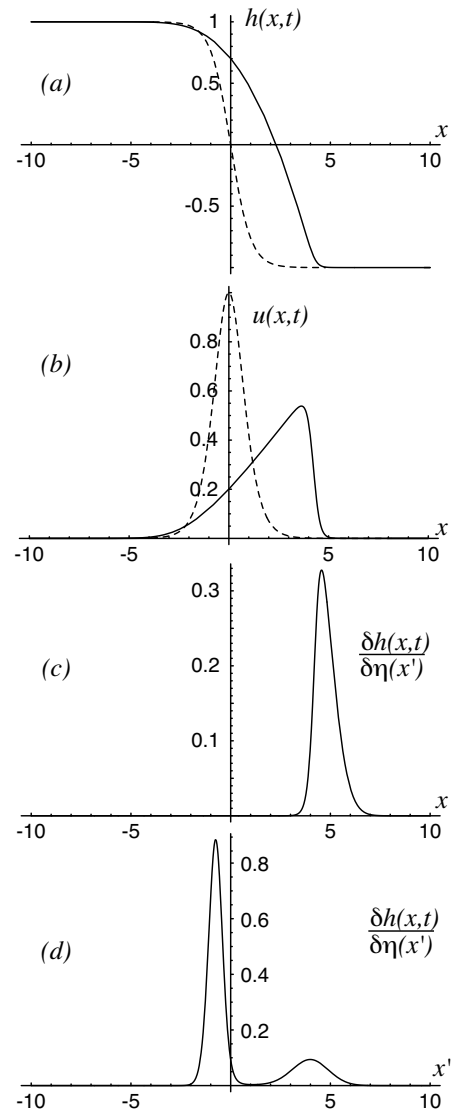}}
     \caption{Plots of the solution to Eq. (\ref{kpza}): (a) and of the 
     solution to the Burgers equation (b) for the time $t=8$ in 
     arbitrary units. The doted lines correspond to initial conditions.
     The sensitivity coefficients are plotted as function of $x$ on 
     (c) and as function of $x'$ on (d).}
      \label{ph1}
\end{figure}

A simple example of illustration of these results may be given with 
the choice of the initial condition for the velocity in the 
Burgers equation as $u(x,0)={\rm{sech}}^{2}\,x$, so that the initial 
condition for $h$ is $h(x,0)=-\tanh\,x$.
On Fig. (\ref{ph1} -- a and b), we present the values of $h(x,t)$ and 
of the solution to the Burgers equation
$u(x,t)$ for the time $t=8$ (given in arbitrary units) as functions of $x$. 
The dotted lines correspond to 
initial values of these functions. The viscosity is fixed to the rather 
small value: $\nu=0.05$.
The sensitivity coefficient $\delta{h(x,t)}/\delta{\eta(x')}$  as a function
of $x$ is given on Fig. (\ref{ph1} -- c) for a fixed value of $x'=4$.  
We observe a maximum value of the 
sensitivity at the bottom of the wave front. This is also the case 
for the sensitivity coefficient as a function of $x'$ where $x$ is 
also fixed to 4. 
However in this case, a higher maximum is found  near the edge of the wave 
front on Fig. (\ref{ph1} -- d). 
This suggests that the whole wave front is involved in the evolution. 
In fact, a better representation of the phenomena is found in the 
$(x,x')$ plane of the sensitivity coefficient. 
\begin{figure}[!ht]
	 \centerline{\includegraphics[scale=0.55]{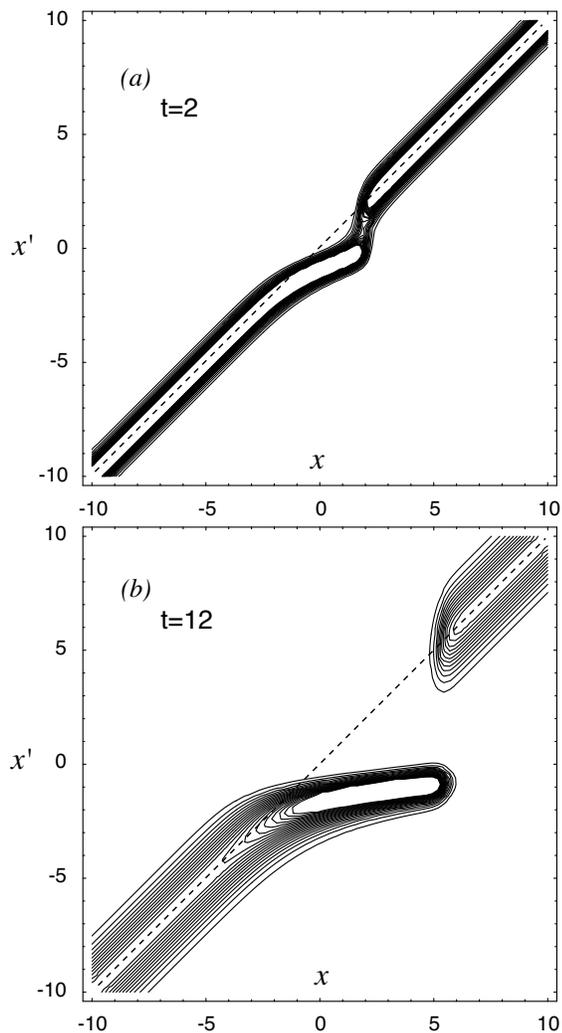}}
     \caption{Contour plot of the sensitivity coefficient
      $\delta{h(x,t)}/\delta{\eta(x')}$ in the $(x,x')$ plane, for two values of 
     time ($t=2$ and $t=12$). For a smaller value of time (a),
      the solution is rather well correlated to the 
     initial condition except in a small region near the wave front. 
     When the time increases (b), the density spreads and one can observe 
     a decorrelation through along the wave front.}
      \label{ph2}
\end{figure}
On Fig. (\ref{ph2}), we 
present the contour plot of the sensitivity coefficient. The doted lines given 
on Fig. (\ref{ph2}) correspond to the initial value of 
the density where the solution is perfectly correlated to the initial 
condition for $\delta{h(x,0)}/\delta{\eta(x')}=\delta(x-x')$.  For a given 
value of $x'$ and coming from large values of $x$ at time $t=2$ (Fig. 
(\ref{ph2}--a)), the sensitivity is 
still well correlated until one reaches a breaking off near 
the bottom of the wave front showing perturbations appear. 
After an interval, larger as the time 
elapses ($t=12$, Fig. (\ref{ph2}--b)), we observe a decorrelation through 
along the wave front. In 
the mean time, the sensitivity coefficient takes its larger values on 
this region. Behind the wave front, the density will be correlated 
again, although we constat a spreading when the time increases. For 
very large value of the time, the density is completely decorrelated as 
we said above with Eq. (\ref{decor}). In order to sum up these 
results, one can say that the most  the solution is sensitive to the
initial condition, the most it is decorrelated in space, \textit{i.e.} all the 
values of the initial condition participate to the solution at any 
distance along the wave front. This finally reveals a characteristic 
of the Burgers turbulence.

Now we can find the sensitivity of the solution to the unforced 
Burgers equation by the same method. In order to do the calculation, we 
have on one hand:
\begin{equation}
    \dfon{u(x,t)}{\varphi(x')}=\int_{-\infty}^{+\infty}
    \dfon{u(x,t)}{\eta(y)}\dfon{\eta(y)}{\varphi(x')}\,\d y
   = \int_{x'}^{+\infty}\dfon{u(x,t)}{\eta(y)}\,\d y,
\end{equation}
for we have 
 $\delta{\eta(y)}/\delta{\varphi(x')}=\theta(y-x')$, where $\theta$ stands for the step 
function and on the other hand:
\begin{equation}
    \dfon{u(x,t)}{\eta(y)}=
    \int_{-\infty}^{+\infty}
    \dfon{u(x,t)}{h(z,t)}\dfon{h(z,t)}{\eta(y)}\,\d 
    z=-\partial_{x}\left[\dfon{h(x,t)}{\eta(y)}\right].
\end{equation}
These results allow us to calculate explicitly the sensitivity of 
the solution to the Burgers equation 
in term of Eq. (\ref{senKPZ}). This is a rather cumbersome relation
although simple to calculated, it is left to the reader. Moreover, 
the main conclusions about 
the sensitivity to initial conditions can  be derived in the same 
manner as the one we have obtained above for $h(x,t)$.

\section{Conclusion}

The question arises now of the possibility to extend the results 
issued from the preceding section to a larger class of equations. 

Let us consider first the forced Burgers equation by a pure time dependent term:
\begin{equation}
    \partial_{t}u+u\partial_{x}u=\nu \partial_{xx}u+f(t).
    \label{inho}
\end{equation}
Orlowski and Sobczyk \cite{os} have found that an appropriate transformation 
onto the variable $x$ and function $u$ maps Eq. (\ref{inho}) to the unforced 
Burgers equation:
\begin{equation}
    y=x-\varphi(t),\,v(y,t)=u(x,t)-\psi(t),
\end{equation}
where
\begin{equation}
   \varphi(t)=\int_{0}^t f(\tau)\,\d\tau\;{\rm{and}}\;
   \psi(t)=\int_{0}^t \varphi(\tau)\,\d\tau.
\end{equation}
Thus the sensitivity coefficient of the $h$ function in this 
transformation just undergoes a translation into Eq. (\ref{senKPZ}):
\begin{equation}
    x\to x-\int_{0}^t f(\tau)\,\d\tau\,
\end{equation}
and does not affect the density otherwise. Then we are brought to the same 
conclusions as in the case of the unforced Burgers equation.

Now we can give few additional remarks. The calculation of the sensitivity 
coefficient of Burgers and related 
equations may be quite naturally extended to the three dimensional 
case because the Hopf-Cole transformation still applies in this case, 
from which the densities may be easily deduced. There is also the case 
of initial/boundary problem for the Burgers equation. Actually, an 
analytic solution to this problem where solved several years ago 
\cite{calo,lilo4,lilo1}, and recently applied to related problems 
concerning the Burgers equation \cite{lilo3,ablo,lilo2}. Sensitivity to 
initial conditions  of such problems seems to be usefull to perform and 
will be the aim of a forthcoming paper.


\end{document}